\documentclass[11pt,a4paper]{nature}
\usepackage{graphicx}

\title{Spectrometry of the Earth using Neutrino Oscillations}

\author{
C. Rott$^1$,
A. Taketa$^2$,
D. Bose$^1$
}

\begin{document}

\maketitle

\begin{affiliations}
\item Department of Physics, Sungkyunkwan University, Suwon 440-746,
Korea.
\item Earthquake Research Institute, University of Tokyo, 1-1-1 Yayoi,
Bunkyo-ku, Tokyo, Japan.
\end{affiliations}
\let\thefootnote\relax\footnote{ Correspondence and requests for
materials should be addressed to C.R. (email: rott@skku.edu) and
A.T. (email: taketa@eri.u-tokyo.ac.jp). Order of first and second
authors is determined by lot. }

\begin{abstract}

The unknown constituents of the interior of our home planet have
provoked the human imagination and driven scientific exploration.  We
herein demonstrate that large neutrino detectors could be used in the
near future to significantly improve our understanding of the Earth's
inner chemical composition. Neutrinos, which are naturally produced in
the atmosphere, traverse the Earth and undergo oscillations that
depend on the Earth's electron density. The Earth's chemical
composition can be determined by combining observations from large
neutrino detectors with seismic measurements of the Earth's matter
density. We present a method that will allow us
to perform a measurement that can distinguish between composition
models of the outer core. We show that the next-generation
large-volume neutrino detectors can provide sufficient sensitivity to
reject outer core models with large hydrogen content and thereby
demonstrate the potential of this novel method. In the future,
dedicated instruments could be capable of distinguishing between
specific Earth composition models and thereby reshape our
understanding of the inner Earth in previously unimagined ways.

\end{abstract}

\newpage

%\section*{Introduction}
%\label{sec:Introduction}
Understanding the inner structure and composition of the Earth is
fundamental to Earth science. While Earth's matter density
distribution can be inferred from geophysical observations, its
compositional structure is far more difficult to determine. The state
and composition of the core, which constitutes 32\% of Earth's mass
and 16\% of its volume, remains largely uncertain. The core consists
of an iron nickel alloy and is divided into inner and outer regions
distinguished by a large density difference at a depth of
approximately 5,100~km. The inner core is solid, while the lack of
s-wave propagation in the outer core and lower density indicate it to
be liquid. The density deficit in the outer core, however, cannot be
simply explained by a difference in state, but rather requires the
presence of light elements at 5 wt\% to 10 wt\%. There is great
excitement in Earth science with regard to determining these light
components in the outer core in order to understand the evolution of
the Earth and the geodynamo.  We introduce a new technique based on
neutrino oscillations in order to remotely measure electron density
and demonstrate how, in the near future, this method could be used to
distinguish between different composition models of the inner Earth.

Analyses of seismic waves have resulted in the well-understood shell
structure of the Earth, consisting of crust, upper mantle, lower
mantle, outer core, and inner core. The matter density structure of the
Earth has been accurately determined by combining astronomic-geodetic
parameters, free oscillation frequencies, and seismic wave velocity
measurements~\cite{Dziewonski1981}. The composition of the crust near
the surface can be measured directly. Drill core samples have resulted
in composition measurements down to a depth of approximately 
12~km~\cite{Popov1999}. The upper mantle composition can be probed
through eruption entrainment sampling~\cite{Hofmann1997}. The state
and composition of the Earth's core, at a depth of approximately 2,900~km
remains far more uncertain with no prospects
of sampling materials.

The outer core composition can be inferred to be mostly iron-nickel
alloy with traces of light elements, by combining seismological
velocity profiles and the composition of primitive
meteorites~\cite{McDonough1995}.  Through recent progress in
high-pressure experiments, hydrogen, carbon, oxygen, silicon, and
sulfur have been suggested as light element
candidates~\cite{Li2007}. However, the abundance of these light
elements remains uncertain.

Obtaining reliable estimates for the abundances of light elements in
the Earth's core is essential to understanding the formation and
evolution of the Earth~\cite{Allegre1995} and to determining the
origin of the geomagnetic field~\cite{Fearn1981}, which are two of the
major unsolved mysteries in Earth science.

Neutrinos (denoted $\nu$) are remarkable particles that have enjoyed
an ever more important role in particle physics, cosmology, and
astrophysics since they were predicted by theorist Wolfgang Pauli in
1930 and first observed in 1956~\cite{Kruse1956}.  There exist three
different types (referred to as flavours) of neutrinos,
$\nu_{e}$, $\nu_{\mu}$, and $\nu_{\tau}$, which relate to how the
neutrino was produced. However, a neutrino's flavour can
change. For example, a neutrino produced as a $\nu_{\mu}$ can be
detected as a $\nu_{e}$. This process, which solved the solar neutrino
problem~\cite{Bahcall1992}, is known as neutrino
oscillation~\cite{Pontecorvo1957}. Neutrino oscillations are a quantum
mechanical consequence of neutrinos having mass, and as such the
behaviour of these oscillations can be described precisely.

In the present study, we propose a novel technique for measuring the
average chemical composition of the deep Earth using neutrinos. Due to
their tiny interaction cross section, neutrinos can pass through the
entire Earth without interacting.  As mentioned earlier, due to
neutrino oscillations, a flavour of one neutrino can covert to another
flavour. Neutrino oscillations depend on the medium traversed, or,
more specifically, on the electron density along the path of the
neutrino through the Earth~\cite{Barger1980}. The compositional
structure of the Earth can be obtained as the average ratio of the
atomic number to the atomic weight (Z/A), by comparing the electron
density distribution and the Earth's matter density distribution. This
effect makes neutrinos unique messenger particles to remotely probe
the Earth's interior.

Large-volume neutrino detectors have emerged as powerful tools in
particle physics and astrophysics. Operating instruments have
demonstrated their tremendous potential in groundbreaking
discoveries, such as the observation of high-energy extra-terrestrial
neutrinos by IceCube and through the observation of neutrino
oscillations by Super-Kamiokande. There is a great interest in
constructing the next generation of neutrino detectors with larger volumes
and improved performance. This new generation of large-volume
detectors could be capable of observing neutrinos at sufficiently high
rates to perform the first experimental measurement of the Earth's
interior.  For example, with the advent of Hyper-Kamiokande
(Hyper-K)~\cite{Abe2011} and the Precision IceCube Next-Generation
Upgrade (PINGU)~\cite{Aartsen}, spectrometry using neutrino
oscillations could enable us to, for the first time, directly
determine the compositional structure of the Earth. Even more
visionary ideas, such as large ocean-going~\cite{Kistler:2008us} or
ice-based detectors, could see neutrino spectrometry emerge as a
precision science.

Preceding research of geophysics using neutrinos can be divided into
three categories: (1) measurement of the radioactive nuclei density in
the Earth using geo-neutrinos generated through nuclear decays, (2)
measurement of Earth's matter density using neutrino absorption, and
(3) measurement of Earth's matter density using neutrino
oscillations~\cite{Jacobsson:2002nb,Lindner2003,Geller2003,Winter2005,Winter:2006vg,Gonzalez-Garcia2008,Agarwalla:2012uj}.
In the present study, we introduce a new fourth category. We apply
neutrino oscillations for a composition measurement, exploiting the
fact that neutrino oscillations are dependent on electron density,
which is the product of the matter density and the ratio of the
average atomic number to the atomic weight. Although the underlying
physical phenomena are well understood, we focus in particular on the
relevance of these effects to geophysics and discuss the prospects for
an Earth composition measurement that could be performed within the
next two decades.

\section*{Results}

\subsection{Neutrino oscillations in the Earth}
\label{subsec:Oscillations}

In geophysics, neutrinos have received attention due to the
information on the inner Earth they provide, as demonstrated by the
measurement of radiogenic heat generated in the Earth through the
observations of neutrinos from nuclear decays of uranium and
thorium~\cite{Araki2005}. The success in detecting these geoneutrinos
has confirmed the feasibility of using neutrinos in Earth
science. While geoneutrinos are generated through nuclear decay and
carry energies of approximately $10^{6}$~eV (one electron volt (eV) $
= 1.602 \times 10^{-19}$~joules), the neutrinos used for the proposed
method have energies of a few GeV~($10^{9}$ eV) and are naturally
produced when energetic cosmic rays collide with the upper Earth's
atmosphere.

The majority of atmospheric neutrinos produced are type $\nu_{\mu}$,
and their flavour changes as they pass straight through the
Earth.  The neutrino oscillation probability depends on a set of
oscillation parameters, the neutrino energy, $E_{\nu}$, the distance
travelled, and the electron density along its path. The path length, L,
is the distance that the neutrino travels from its point of origin in
the atmosphere to the detector.  Since all neutrinos relevant for this
analysis are generated in the Earth's atmosphere, L is simply a
function of the zenith angle, $\Theta$, of the neutrino arrival
direction at the detector. Figure~1(a) shows the neutrino path through
the Earth.

We calculate neutrino oscillation probabilities, following the
approach of Barger et al.~\cite{Barger1980} and use the numerical
implementation of the NuCraft software
package~\cite{Wallraff2014}. The oscillation parameters, which are
well measured, are taken from the global fit given by Capozzi et
al.~\cite{Capozzi2014}, assuming the case of a normal mass hierarchy,
as favoured in current measurements. We use the modified Preliminary
Reference Earth Model (PREM) matter density
model~\cite{Dziewonski1981,Durek1996} to describe the Earth density
and structure. We fix the mantle composition to pyrolite and the inner
core composition to iron, only the outer core composition is
varied. Figure~1(b) shows the $\nu_{\mu}$ survival probability and the
$\nu_{e}$ appearance probability as a function of the path length for
a neutrino with an energy of 4~GeV~($10^{9}$~eV) passing vertically
through the Earth. The survival probability is the probability that a
created neutrino of specific flavour is observed as such. In
this case, we consider a muon neutrino observed as such $P(\nu_\mu
\rightarrow \nu_\mu)$. The appearance probability is the chance that a
neutrino of one flavour is observed as a neutrino of a
different flavour, for example $P(\nu_\mu \rightarrow \nu_e)$.
The flavour change as a function of travelled distance in the
Earth is shown. In order to visually show the impact of the outer core
composition on the oscillation probability, we compare the cases of an
alloy of iron and 2 wt\% (weight percent) hydrogen with
iron. Figure~1(c) shows the $\nu_{\mu}$ survival probability at the
surface of the Earth, as a function of the neutrino's energy for four
different core compositions. In order to visualize the
difference in survival probability for different outer core
compositions, we selected (1) iron, (2) an alloy of iron and 1 wt\%
hydrogen, (3) an alloy of iron and 2 wt\% hydrogen, and (4) an alloy
of iron and 5 wt\% hydrogen as extreme examples of the outer core
composition.

%%%Figure 1
\begin{figure}[htbp]
\centering
\includegraphics[width=0.98\textwidth]{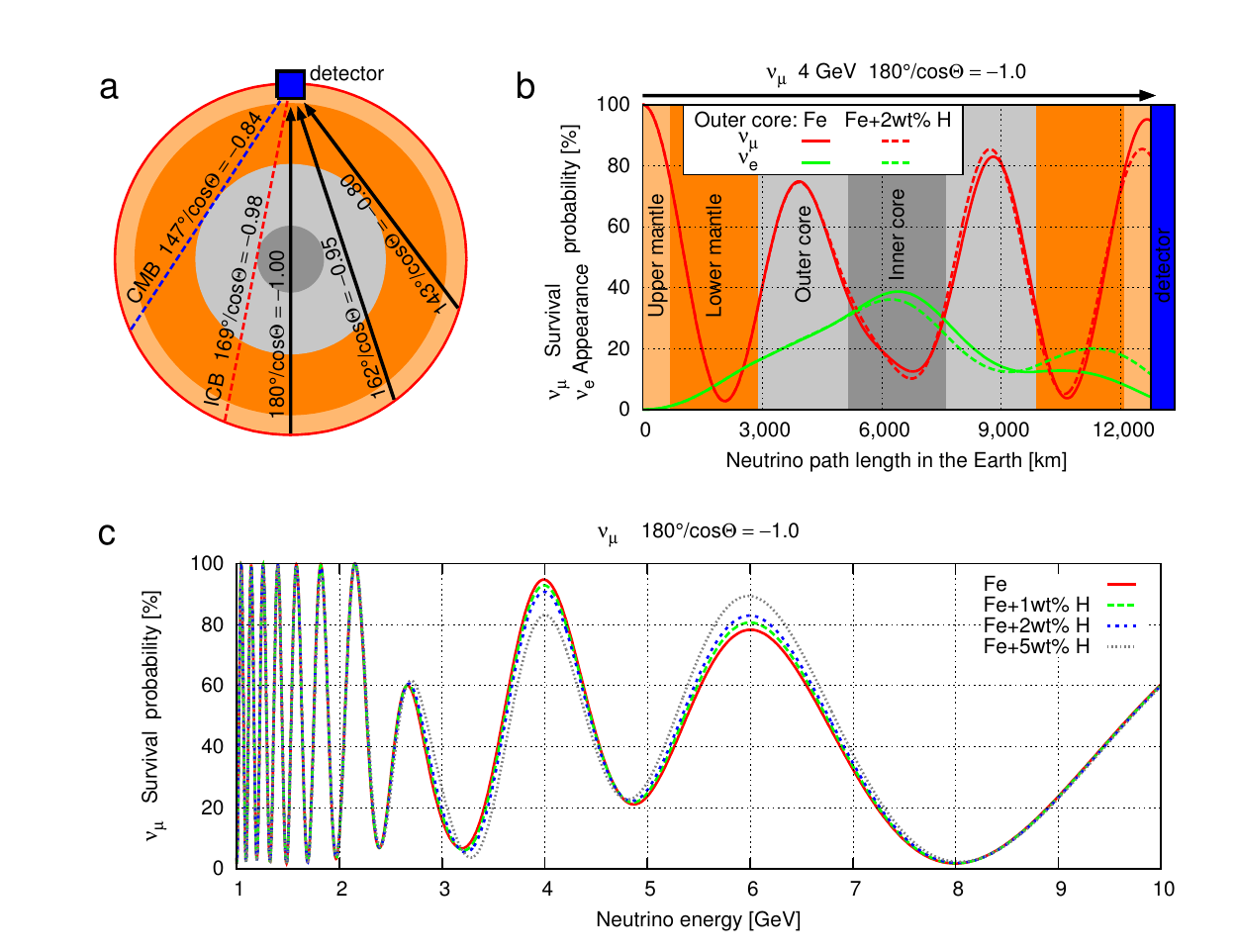}
\caption{ (a) Schematic diagram of a neutrino's path through the Earth
  and the corresponding zenith angles. The inner core boundary (ICB)
  at $\Theta=169^{\circ}$ and the core mantle boundary (CMB) at
  $\Theta=147^{\circ}$ are indicated by dashed red and blue lines,
  respectively. (b) $\nu_{e}$ appearance probability (green) and
  $\nu_\mu$ survival probability (red) as functions of path length in
  the Earth. The neutrino direction is $\Theta=180^{\circ}$, as shown
  in (a). The solid/dashed line corresponds to the case in which the
  composition of the outer core is pure iron/an alloy of iron and 2
  wt\% hydrogen. (c) $\Theta=180^{\circ}$-$\nu_\mu$ survival
  probabilities as a function of neutrino energy for different outer
  core compositions. The solid~(red), long dashed~(green), short
  dashed~(blue), and dotted~(gray) lines represent iron, an alloy of
  iron and 1 wt\% hydrogen, an alloy of iron and 2 wt\% hydrogen, and
  an alloy of iron and 5 wt\% hydrogen, respectively.}
  \label{fig1}
\end{figure}

\subsection{Z/A ratios for different outer core models}
\label{subsec:zaratios}
Iron is the most abundant element in the outer Earth core and
throughout this document we have chosen pure iron as our default
composition. Models adding single or multiple elements to iron have
been proposed~\cite{Allegre2001,Mcdonough2003,Huang2011}.  In Table~1,
we introduce some selected outer core composition models and
characterize them according to Z/A ratio. The estimated maximal
abundance of light elements~\cite{Poirier1993,Li2007} for alloys of
iron are listed in Table 1. Note that nickel is thought to co-exist
with iron in the outer core, with an estimated content of
approximately 5\%~\cite{Wood2006}. Since there is only a slight
difference between Z/A values, using an alloy of iron and 5 wt\%
nickel as the base composition rather than iron will result in only a
marginal change in Z/A from 0.4656 to 0.4661.

\begin{table}[htbp]
\caption{Z/A ratios for alloys of iron and light elements and some selected composition models.}
\centering
\begin{tabular}{|l||c|c|c|c|c|c|l|}
\hline
Model name & Z/A ratio & Si(wt\%) & O(wt\%) & S(wt\%) & C(wt\%) & H(wt\%) & reference  \\
\hline
\hline
\multicolumn{8}{|l|}{Single-light-element model (maximum abundance)} \\
\hline
Fe+18wt\%Si & 0.4715 & 18 & -  & -  & -  & - & Poirier~\cite{Poirier1993} \\
\hline
Fe+11wt\%O  & 0.4693 & -  & 11 & -  & -  & - & Poirier~\cite{Poirier1993} \\
\hline
Fe+13wt\%S  & 0.4699 & -  & -  & 13 & -  & - & Li and Fei~\cite{Li2007} \\
\hline
Fe+12wt\%C  & 0.4697 & -  & -  & -  & 12 & - & Li and Fei~\cite{Li2007} \\
\hline
Fe+1wt\%H   & 0.4709 & -  & -  & -  & -  & 1 & Li and Fei~\cite{Li2007} \\
\hline
\hline
\multicolumn{8}{|l|}{Multiple-light-element model} \\
\hline
Allegre2001   & 0.4699 & 7  & 5  & 1.21 & -  & - & All\`{e}gre et al.~\cite{Allegre2001} \\
\hline
McDonough2003 & 0.4682 & 6  & 0  & 1.9 & 0.2 & 0.06 & McDonough~\cite{Mcdonough2003} \\
\hline
Huang2011 & 0.4678 & -  & 0.1  & 5.7 & - & - & Huang et al.~\cite{Huang2011} \\
\hline
\end{tabular}
\end{table}

\subsection{Oscillation probabilities for different outer core models}
\label{subsec:Oscillogram}

As neutrino oscillations simply depend on the neutrino's energy, path
length, and composition along the path, we can determine the
probability that a neutrino will change flavours as a function of the
zenith angle and the energy. We calculated the oscillation
probabilities for different core models.  Figure~2 shows such an
oscillogram, i.e., oscillation probabilities as a function of zenith
angle and neutrino energy, for two different outer core
compositions. Subtle differences in the neutrino survival probability
can be exploited in order to distinguish between different composition
models. The most pronounced differences in survival probability are
for neutrinos with energies between 2~GeV and 8~GeV that traverse the
outer core, i.e., their zenith angles are larger than $147^{\circ}$.

\subsection{Detector requirements}
The described differences in neutrino oscillation effects that depend
on the Earth's composition could be detectable with a neutrino
detector if the detector combines good energy and angular resolution
in the relevant energy range and observes GeV neutrinos at
sufficiently high rates to accumulate sufficient statistical
samples. Due to the small neutrino interaction cross section, a large
detector volume of megaton scale is necessary in order to acquire a
sufficient number of neutrino events and not suffer from limited
statistics. Good neutrino flavour identification can be
beneficial.

\subsection{Neutrino detectors}
\label{subsec:Detectors}
Large neutrino detectors have been realized in a cost-effective manner
by using water or ice as a naturally occurring detector medium to
observe Cherenkov light emissions from one or more energetic particles
produced in neutrino interactions.  The IceCube neutrino
telescope~\cite{Ahrens2004} uses one gigaton (1,000 megatons) of ice
at the Geographic South Pole that was instrumented with more than
5,000~photosensors.  The optical sensor array relies on the ultra-pure
Antarctic ice as a detection medium. The detector is working extremely
well, and the recent discovery of high-energy astrophysical neutrinos
demonstrates the potential of large neutrino
detectors~\cite{Aartsen2013}. Super-Kamiokande~\cite{Fukuda1998},
which is the leading high-precision detectors, has a 50~kiloton water
tank surrounded by 11,000~photosensors to observe Cherenkov light,
allowing neutrino energies to be determined with high precision and
neutrino flavours to be identified reliably. The underlying technology
applied by IceCube and Super-Kamiokande is well established and is the
basis for future detectors. Next-generation detectors could benefit
from better photosensors with higher photon detection efficiency.

%%%Figure 2
\begin{figure}[htbp]
  \begin{center}
    \includegraphics[width=0.98\textwidth]{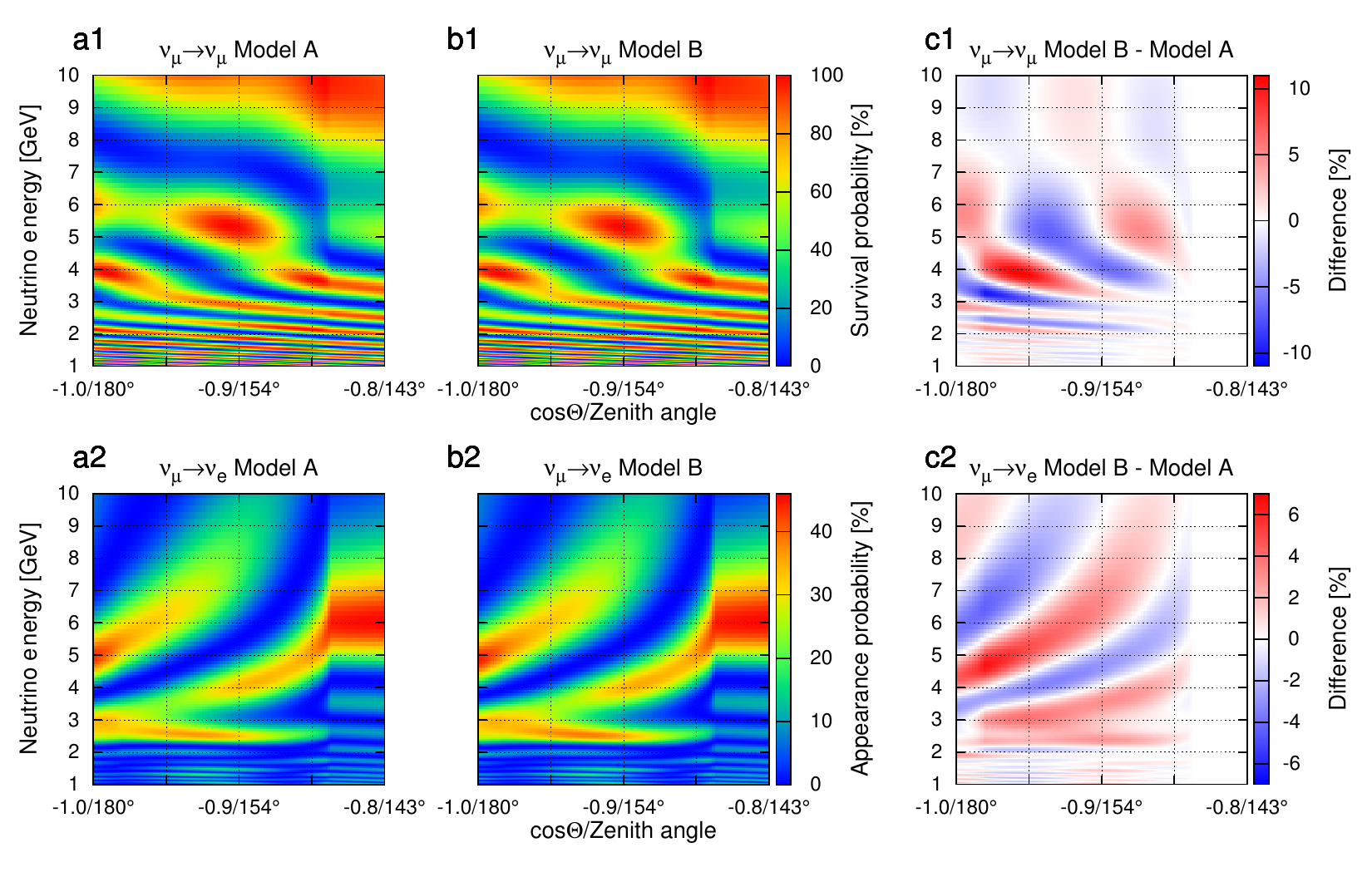}
\caption{Comparison of oscillation probabilities for two different
  core compositions: Model A -- iron; Model B -- an alloy of iron and
  2 wt\% hydrogen. a1 and a2 show the $\nu_\mu$ survival probabilities
  as a function of neutrino energy and zenith angle for Models A and
B, respectively. b1 and b2 show the appearance probability for
  $\nu_{\mu}$ to $\nu_{e}$ for Models A and B, respectively. a3
  shows the difference between a1 and a2, and b3 shows the difference
  between b1 and b2.}
  \end{center}
  \label{fig2}
\end{figure}

\subsection{Sensitivity of benchmark detectors}
\label{subsec:Benchmark}

We calculate the confidence level with which the composition of the
outer core could be determined for some benchmark neutrino
detectors. We use a generic neutrino detector description based on
performance parameters to estimate sensitivities. Our parameterization
can easily be converted into hardware and design requirements for the
planning of new detectors. We compute expected event rates as a
function of the neutrino energy and the zenith angle as a function of
the product of detector size and the exposure time in megaton-years.
In this way, we calculate the number of neutrino events for a certain
energy, direction, and flavour. We create templates of the expected
event rates for different outer core models. Event rates were
calculated from the atmospheric neutrino flux, oscillation
probabilities, neutrino cross section, detector volume, and exposure
time. The atmospheric neutrino flux and energy spectrum are well
understood for our purposes. We adopt the atmospheric neutrino
flux model of Athar et al.~\cite{SajjadAthar2013}.  For the neutrino
($\nu_\mu$) and anti-neutrino ($\bar{\nu}_\mu$) interaction cross
sections, we use the approximate values of $7.0 \times ({\rm E/GeV})
\times 10^{-39} {\rm cm}^2$ and $3.0 \times ({\rm E/GeV}) \times
10^{-39} {\rm cm}^2$, respectively~\cite{Formaggio2012}.

The outcome of any experimental measurement that deals with individual
events, such as the detection of neutrinos, will be subject to
statistical fluctuations. We consider a large number of potential
experimental outcomes, called pseudo experiments, in order to estimate
the chance that models could be distinguished through an actual
measurement. For each pseudo experiment, we compare the number of
observed events to the number of expected events for a specific
model. We calculate events for given ranges of energy and zenith
angle. For each of these bins in energy and zenith angle, events
follow Poisson statistics and we determine the probability for the
observation. We then compute the likelihood of this experimental
outcome with respect to a specific Earth model assumption. The total
likelihood is then given by the product of the likelihoods for the
individual bins. In order to compare the likelihood of one model with
that of another, we compute the likelihood ratio. For the calculation
of the expected significance, we perform a set of pseudo experiments
and apply the log-likelihood ratio (LLR)
method~\cite{james2006statistical}.

For simplicity of the analysis, we consider only muon neutrino
($\nu_{\mu} + \bar{\nu}_\mu$) events, which are the most relevant for
neutrino spectrometry.  Muon neutrino events can be identified with
high efficiency by proposed next-generation detectors, such as PINGU
or Hyper-K. For PINGU, the resolution for the muon neutrino energy is
expected to be better than 25\% at 5~GeV, and the zenith angle
resolution for the neutrino has been reported to be approximately
$13^{\circ}$~\cite{Aartsen}. At Hyper-K, better angular and energy
resolutions compared to PINGU can be expected, in addition to a larger
than 99\% efficiency to identify the interaction products of a
neutrino interaction~\cite{Abe2011,Abe:2014oxa}.

Figure~3 shows the sensitivity for rejecting outer core compositions
given by their Z/A ratios with respect to iron for a generic neutrino
detector.  The detector is characterized by energy resolution and
angular resolution, as defined by $\Delta E_{\nu} / E_{\nu} = \alpha$
and $\Delta \Theta = \beta / \sqrt{E/GeV} $, respectively. We choose
$\alpha = 0.20$ and $\beta=0.25$ for our default benchmark detector
and show the sensitivity depending on the product of lifetime and
detector volume (megaton-years) in Figure~3(b). With an acquired
dataset of 10~megaton-years, a neutrino detector could for the first
time confirm an iron-like core through experimental measurements.
Outer core compositions dominated by lead or water could be rejected
with more than 99\% confidence with respect to iron. The Z/A ratios of
iron, lead, mantle (pyrolite), and water are 0.4656, 0.3958, 0.4957,
and 0.5556, respectively. The large hydrogen content appearing at the
axis at the top of each plot could be rejected.  Figure~3(a) shows the
prospects of neutrino spectrometry. A one-gigaton-year (or
1,000~megaton-year) dataset, which is equivalent to observation for 20
year using a 50~megaton detector, would provide the ability to
discriminate between established outer core composition
models. Furthermore, the hydrogen content of the outer core could be
measured with a precision of 0.4 wt\%.  Better sensitivities could be
achieved if the detector exceeds the benchmark detector performance
parameters selected for use in the present study. Figures~3(c) and
3(d) show the energy resolution and angular resolution dependences of
the sensitivity, respectively.

%%%Figure 3
\begin{figure}[htbp]
  \begin{center}
    \includegraphics[width=0.75\textwidth]{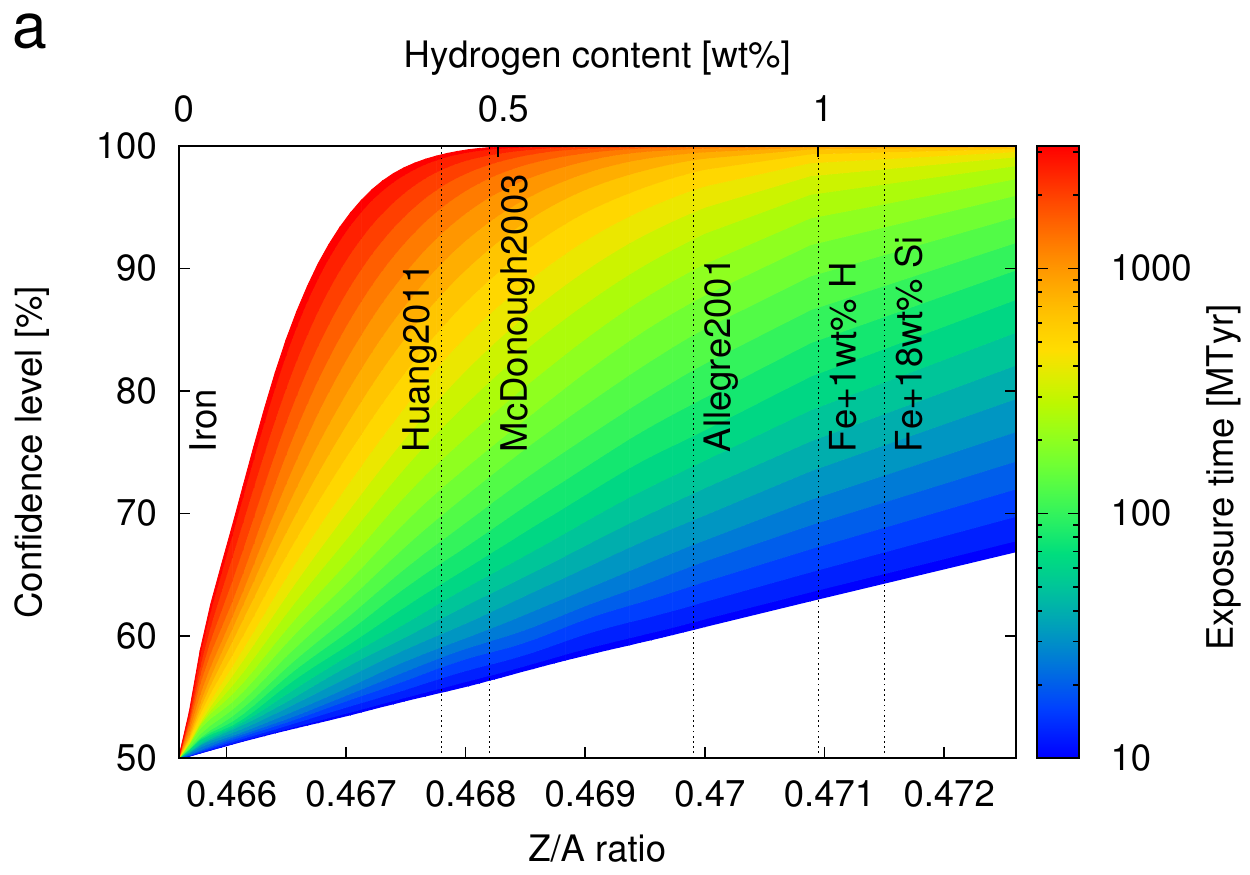}
     \ \\
    \includegraphics[width=0.85\textwidth]{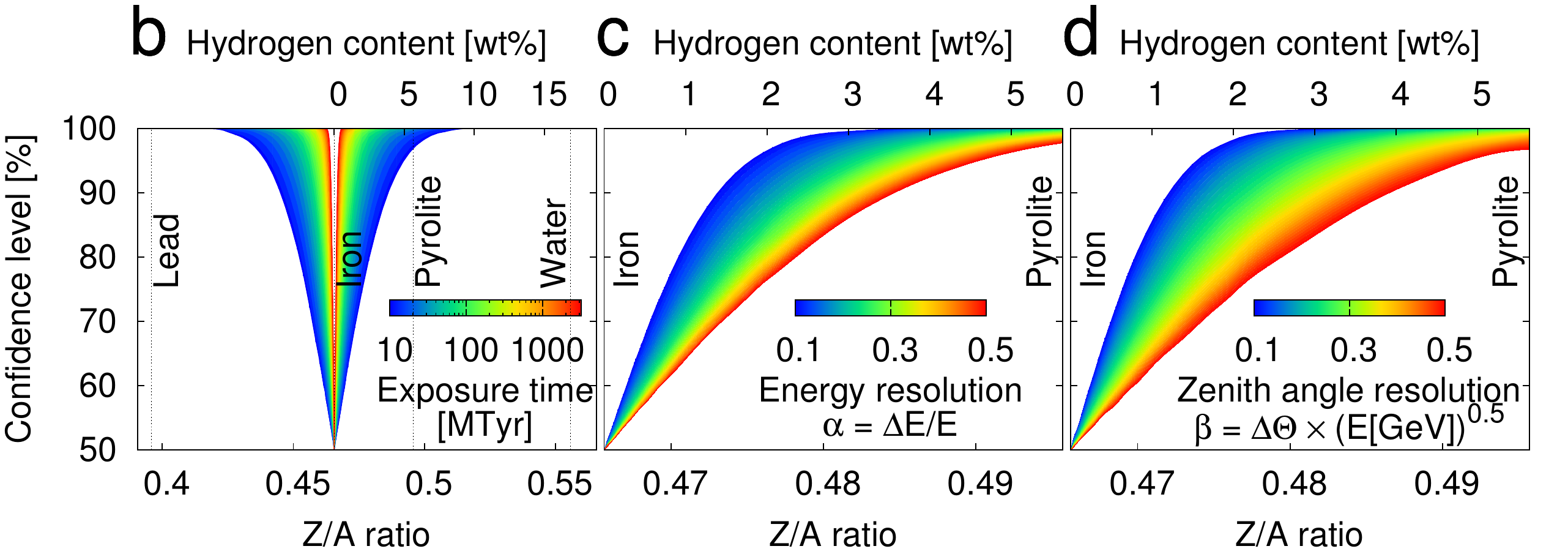}
\caption{(a) Expected confidence level for rejecting a specific outer
  core composition with respect to iron plotted as a function of the
  corresponding Z/A ratio. A generic detector case with an energy
  resolution of 20\% and an angular resolution of $0.25 \times
  (\rm{E/GeV})^{-0.5}$ is shown as an example. The colour indicates
  the exposure time given in megaton-years. We indicate the Z/A ratios
  for some selected outer core composition models (see Table~1 for
  details) as black dotted vertical lines. (b) The same plot as (a)
  for a larger Z/A range. Sensitivity dependences on (c) energy
  resolution and (d) angular resolution for a generic detector with an
  exposure time of 30 megaton-years for an angular resolution of $0.25
  \times (\rm{E/GeV})^{-0.5}$ and an energy resolution of 20\%,
  respectively.}
  \end{center}
  \label{fig3}
\end{figure}

Note that neutrino spectrometry itself only determines the Z/A ratio;
hence, the resulting measurement could be degenerate in corresponding
Earth composition models. A model with Fe~(90 wt\%)+O~(10 wt\%)
(Z/A=0.4690) could be clearly distinguished from an iron core (Z/A =
0.4656), but would have a relatively similar signature as a model with
Fe~(90 wt\%) + Ni~(5 wt\%) + Si~(5 wt\%) (Z/A=0.4681).  Since oxygen,
sulfur, silicon, and carbon have relatively similar Z/A ratios,
neutrino spectrometry would be accompanied by ambiguities in the
measurement of these elements.  High-pressure
experiments~\cite{Hirose2013} combined with neutrino spectrometry
could resolve the remaining degeneracies to allow estimation of the
relative abundances of light elements.

\subsection{Uncertainties}
\label{subsec:Systematic}

We examine our results and the feasibility of the neutrino
spectrometry measurement with future neutrino detectors with respect
to theoretical and experimental uncertainties.

Uncertainties in the composition measurement originate from limited
knowledge of the neutrino oscillation parameters, atmospheric neutrino
flux uncertainties, neutrino cross-section uncertainties, and
uncertainties in the Earth's matter density profile. In addition to
these theoretical uncertainties, detector acceptance-related
uncertainties must be determined. However, this is beyond the scope of
the present study and would have to be carried out through
experimental collaborations.

At present, the limited knowledge of neutrino mixing parameters is the
major source of uncertainty in the proposed neutrino spectrometry
measurement. Using the current best-fit oscillation parameters and
their uncertainty~\cite{Capozzi2014}, the error in the confidence
level curve was approximately $\pm$4\% at 90\% (see Supplementary
Figure~1).  Several experiments are planned for the near future in
order to realize more precise measurements of the neutrino mixing
parameters~\cite{Abe2011a,An2012,Li2014}. A better determination of
the neutrino oscillation parameters will reduce uncertainties.

In order to estimate the uncertainty resulting from the matter density
models, we calculated the confidence level curves using three
different density models (PREM500~\cite{Dziewonski1981,Durek1996},
AK135~\cite{kennett1995}, PEM-A~\cite{Dziewonski1975}; see
Supplementary Figure~2).  The systematic error resulting from the
matter density in the confidence level curve was negligible compared
with the systematic error resulting from the mixing parameters (see
Supplementary Figure~3). The expected uncertainty is sufficiently
small to distinguish the models introduced in the present
study. However, in order to determine the light material contents in
the outer core, a more precise mixing parameter and matter density
model, which may be available in the near future, are needed.

The uncertainty in the atmospheric neutrino flux is estimated to be
approximately 10\%~\cite{Honda2011,SajjadAthar2013}. However, this
uncertainty does not directly affect the Earth composition
measurement, because the atmospheric neutrino flux can be measured by
the neutrino detector itself. The flux ratio
$(\nu_\mu+\bar{\nu}_\mu)/(\nu_e+\bar{\nu}_e)$, on the other hand, may
be affected if the detector cannot reliably provide a particle
identification to distinguish between $\nu_e$ and $\nu_\mu $
events. Since the downward-moving neutrino flux is not subject to
oscillations (due to the short distance), it can be used to measure
the flavour ratio using the detector itself.

At present, the neutrino mass hierarchy, one of the remaining
fundamental neutrino properties, is unknown. By the time that the
proposed composition measurement is performed, we can safely assume
that the mass hierarchy will have been determined, potentially even at
the same neutrino detector considered for our
measurement~\cite{Akhmedov:2012ah,Ge:2013zua}. A normal mass hierarchy
will increase neutrino oscillation probabilities and thus make
neutrino spectroscopy measurements easier. A normal mass hierarchy is
currently favoured~\cite{Fogli2012} in global fits and is therefore
used in the present study.  If the mass hierarchy turns out to be
inverted, the expected neutrino oscillation probabilities at the
neutrino detector will be reduced, and in order to obtain the same
sensitivity as in the normal hierarchy case, a detector would have to
acquire a dataset that is six times larger.

\section*{Discussions}
\label{sec:Discussions}

Neutrino oscillations provide a way to distinguish different Earth
composition models by probing the Z/A ratio. We expect neutrino
spectroscopy to develop into a unique method for measuring the
chemical composition of the inner Earth, which will lead us to a
better understanding of the Earth's evolution and the origin of the
geomagnetic field. Large-volume neutrino detectors that can accumulate
atmospheric neutrino samples at significant rates and have good
angular and energy resolutions at neutrino energies of 2-8~GeV are
needed for these measurements. In the future, new neutrino telescopes
and upgrades to existing instruments will significantly enhance
neutrino detection capabilities in the most relevant energy range for
the spectroscopic measurement discussed here. The Hyper-K project will
see the construction of a 0.6-megaton fiducial volume detector
comprising eight compartments and approximately 100,000~photosensors.
PINGU will use a few megatons of ice in the centre of the IceCube
detector where the ice is clearest.  A detector of similar size is
also considered as a deep-sea neutrino telescope in the Mediterranean
Sea as part of the KM3NeT
project~\cite{katz2009status,katz2014}. These next-generation neutrino
detectors could already offer sufficient sensitivity in order to
exclude extreme models of the Earth's composition. If they are
carefully optimized for neutrino spectrometry, the first meaningful
bounds on the hydrogen content in the core could be in reach.  In the
future, dedicated neutrino experiments could be used to distinguish
between different composition models, as we have demonstrated. We
limited our research to a conservative scenario considering that only
muon neutrinos are detected. The detection of neutrinos of different
flavours is expected to enhance the sensitivity of the proposed
method, warranting further investigation in the future.

In the present study, we focused on the composition of the outer core,
but neutrino spectrometry could also be applied to the mantle,
especially in order to elucidate the water content of the lower
mantle. Through recent progress in diamond inclusion sampling,
high-pressure experiments, and dense seismic velocity measurements, it
was found that the uppermost part of the lower mantle can reserve
1~wt\% water~\cite{Schmandt2014}.  Neutrino spectrometry has the
ability to provide an upper limit for the water content of the lower
mantle in the same way as the hydrogen content of the outer core.

\section*{Methods}
\label{sec:Methods}

\subsection{Flux calculation}

Event distributions for a generic neutrino detector defined by volume,
energy resolution, and angular resolution were calculated for
angle-averaged atmospheric neutrino fluxes after propagation through
the Earth. The calculation proceeded in two steps. We first calculated
the transition probabilities for neutrinos of all flavours as function
of zenith angle and energy. The results were binned in 720 bins of the
cosine of zenith angle and 400 bins of energy to form a transfer
matrix $M$. Full three-flavour neutrino oscillations were performed,
and an Earth structural model with 400 layers (see Supplementary
Figure~2) and various composition models was used. The predicted
neutrino flux for muon neutrinos obtained from the Honda model was
propagated through the Earth using transition tables describing the
muon neutrino survival probability and binned in 40 bins of the log of
reconstructed neutrino energy and in 20 bins of the cosine of the
reconstructed zenith angle for the range of cos$\Theta = (-1...0)$ and
log(E)=$0 ... 1$. The reconstructed zenith angle and neutrino energies
were randomly sampled from the expected distribution as defined by the
detector model (assuming Gaussian distributions). For each bin of $M$,
100,000 events were generated and mapped into the reconstruction
matrix, weighted by the expected Honda flux. Rates were calculated
according to the neutrino interaction cross section and the product of
the detector volume and operation time (megaton-years). Each bin
$m_{ij}$ of $M$ obtains the expectation value of an observation, were
a measurement to be performed. Using a different Earth composition
model and the same generic detector model, we can obtain the
expectation values for this model $m^{*}_{ij}$. Here, $m_{ij}$ and
$m^{*}_{ij}$ act as templates for our log-likelihood analysis in order
to determine the model sensitivity.

\subsection{Log-likelihood method}

Ensembles of pseudo datasets are drawn from each template $m^{*}_{ij}$
and $m_{ij}$ that are repeatedly varied following Poisson
statistics. The log of the Poisson likelihood of the pseudo data for a
specific bin is calculated with respect to the corresponding bin in
$m^{*}_{ij}$ and $m_{ij}$. We take the sum of (-2)
log(P($o_{ij},m_{ij}$)) to obtain the total likelihood. In this way,
for each pseudo dataset, two likelihoods are calculated and are
labelled $\mathcal{L}$( pseudo data, template). The likelihoods are
used to calculate LLR. We calculate two distributions for pseudo data
$o_{ij}$ drawn from $m_{ij}$ given by $\mathcal{L}(o(m)|m)$ /
$\mathcal{L}(o(m)|m^{*})$ and $\mathcal{L}(o(m^{*})|m)$ /
$\mathcal{L}(o(m^{*})|m^{*})$. A total of 10,000 pseudo datasets are
used to achieve adequate coverage of the probability space. We obtain
expected significances from our ensemble of pseudo experiments. The
probability of distinguishing model $m$ from $m^{*}$ is obtained by
calculating the fraction of cases in which events drawn from $m$ have
a likelihood ratio that is more consistent with $m$ than $m^{*}$.

\section*{Acknowledgments}
We would like to thank William McDonough, Kotoyo Hoshina, and Hiroyuki
Tanaka for engaging in useful discussions. The present study was
supported by the Faculty Research Fund, Sungkyunkwan University,
2013. We used the computer systems of the Earthquake and Volcano
Information Center of the Earthquake Research Institute, University of
Tokyo.

\section*{Supplementary Figures}

\renewcommand{\figurename}{Supplementary Figure}
\setcounter{figure}{0}

\begin{figure}[htbp]
\centering
\includegraphics[width=0.7\textwidth]{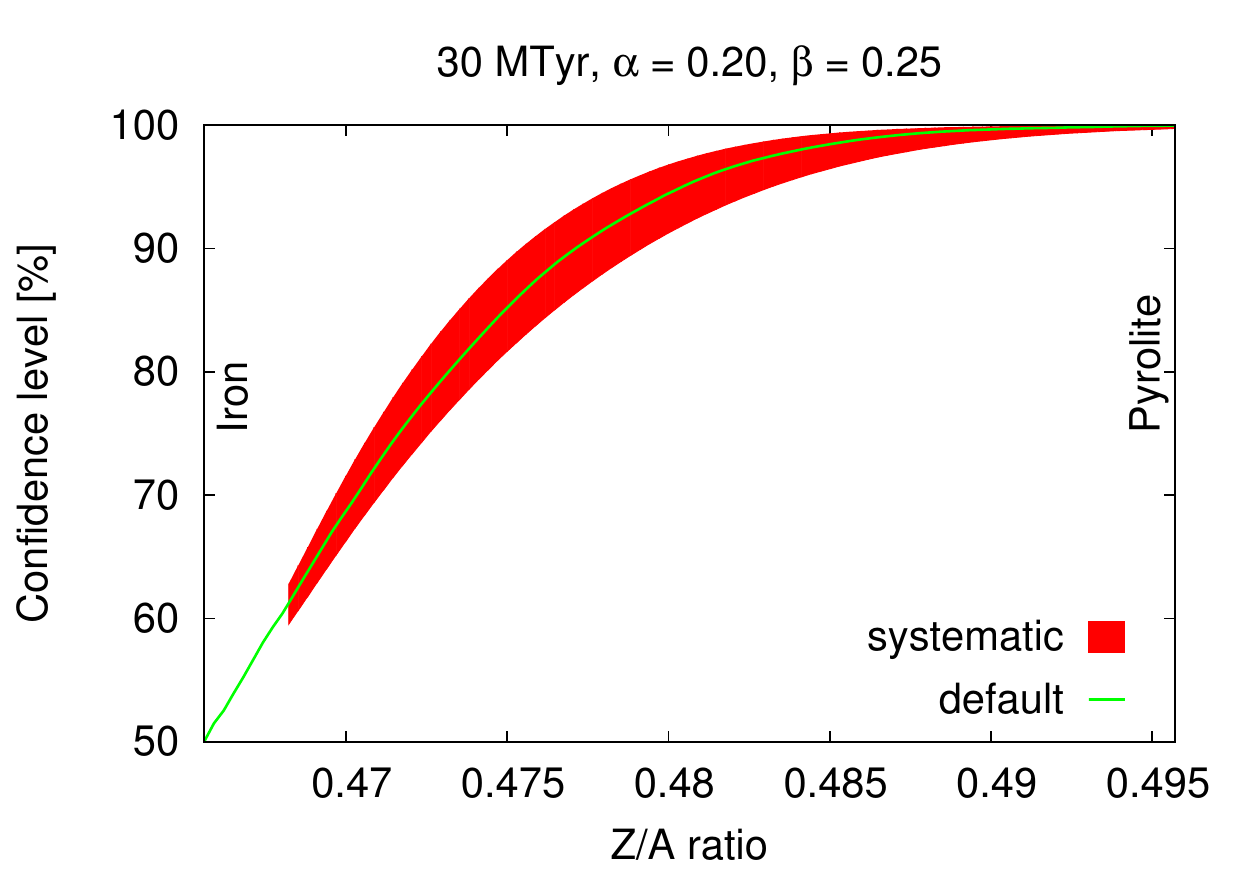}
\caption{Systematic error of the expected confidence level as a
  function of Z/A. The dashed~(green) line represents default mixing
  parameter case and the red area represents its uncertainty. A
  generic detector case with an exposure time of 30 MTyr, an energy
  resolution of 20\%, and an angular resolution of $0.25 \times
  (\rm{E/GeV})^{-0.5}$ is shown.}
  \label{supfig1}
\end{figure}

\begin{figure}[htbp]
\centering
\includegraphics[width=0.49\textwidth]{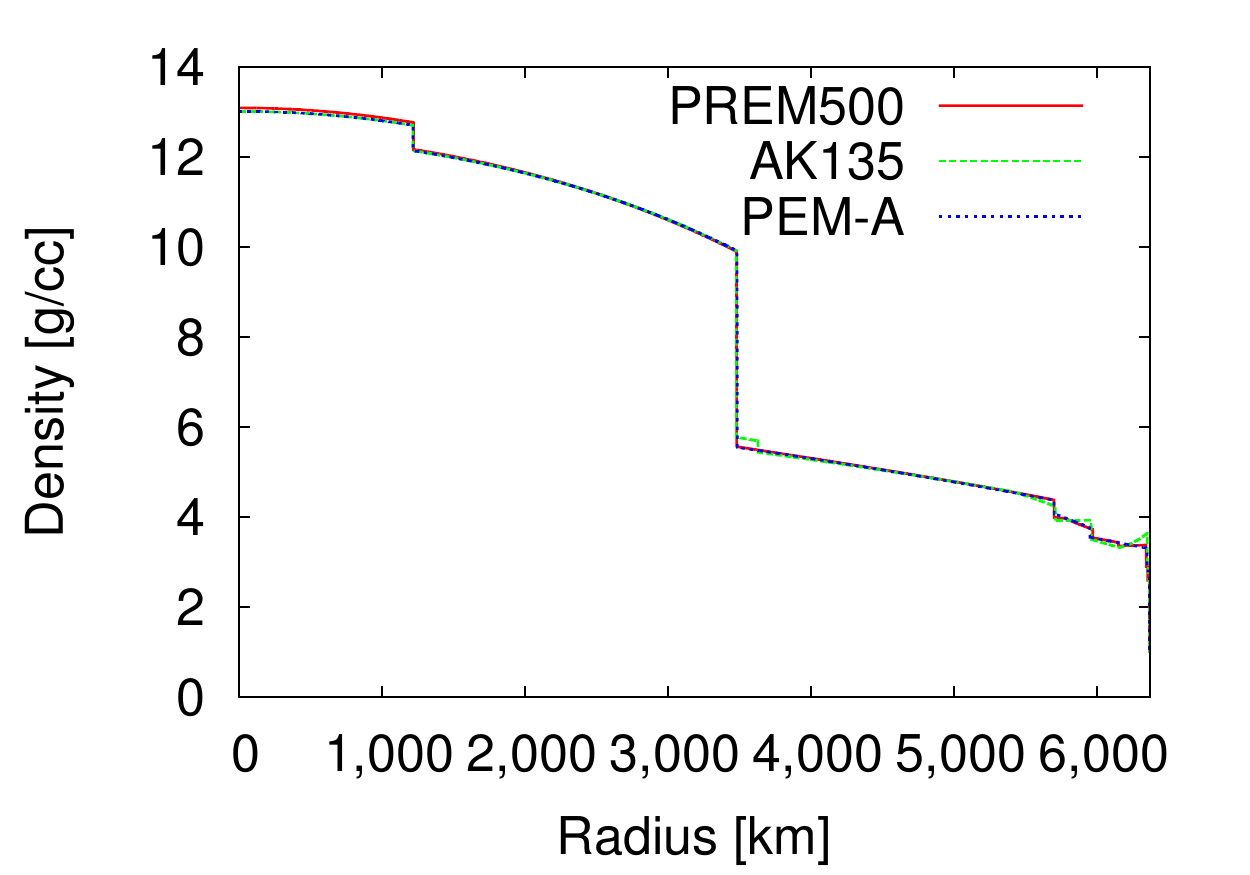}
\includegraphics[width=0.49\textwidth]{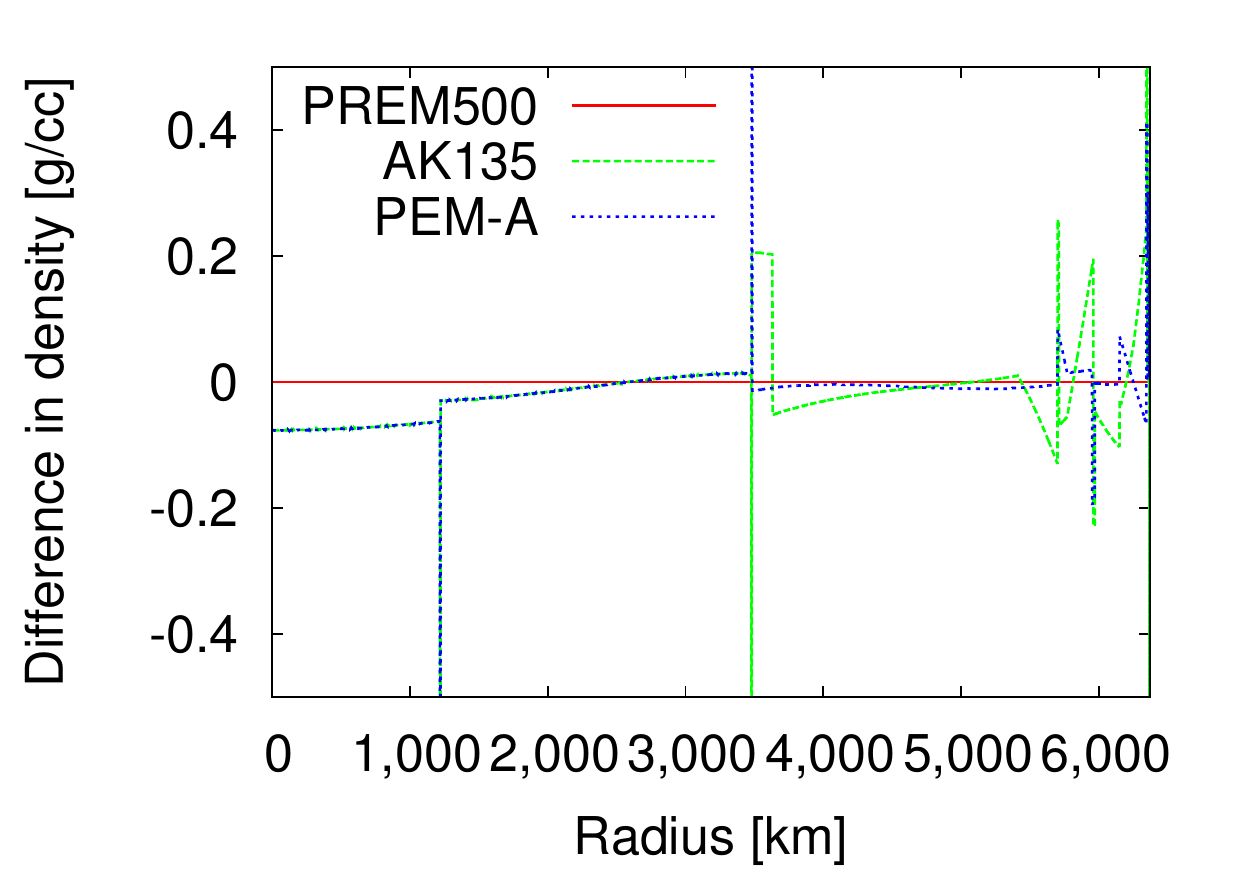}
\caption{left: Matter density distributions of the Earth as a
  function of radius from the centre of the Earth. The solid~(red),
  dashed~(green), and dotted~(blue) lines represent the modified PREM,
  AK135, and PEM-A, respectively. right: Difference from modified PREM
  (PREM500).}
  \label{supfig2}
\end{figure}

\begin{figure}[htbp]
\centering
\includegraphics[width=0.7\textwidth]{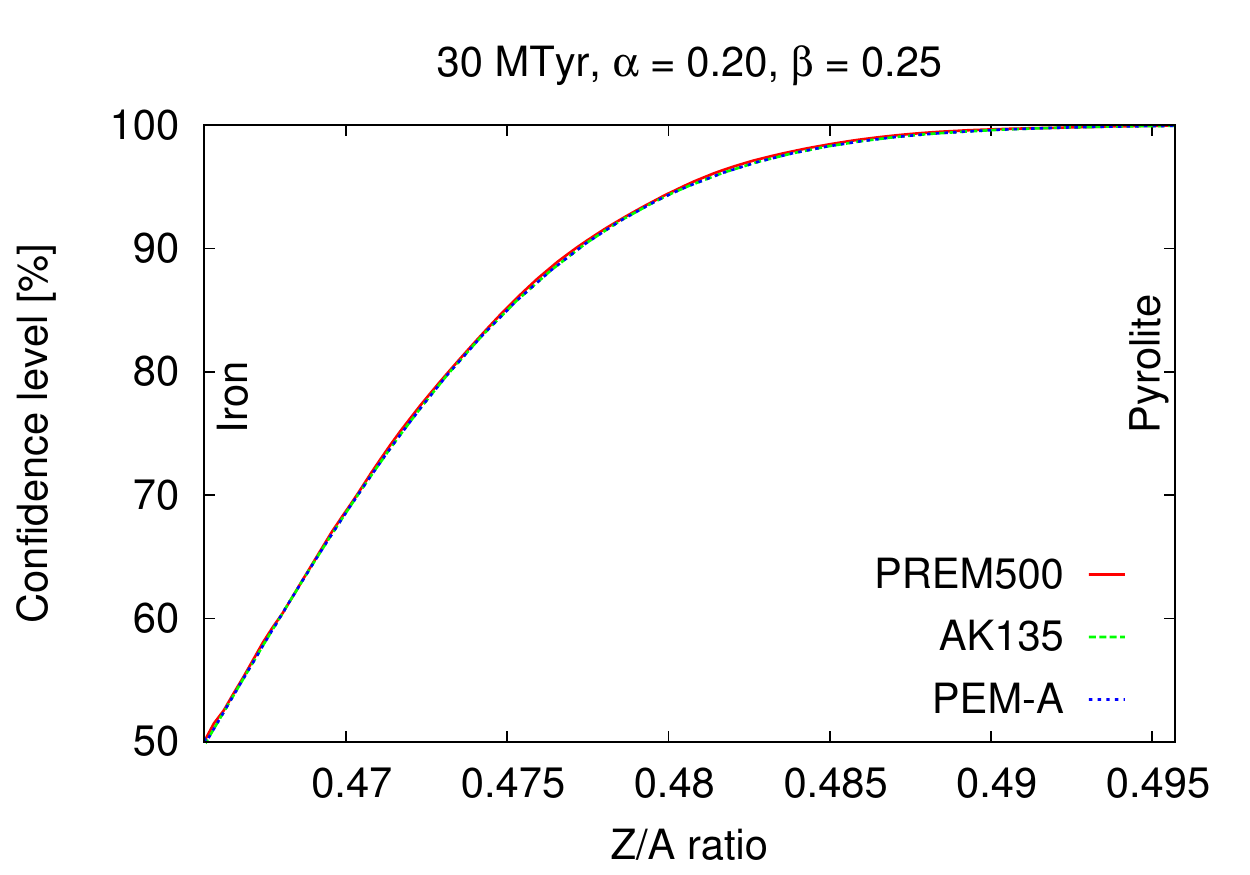}
\caption{Expected confidence level for rejecting a specific outer core
  composition with respect to iron plotted as a function of the
  corresponding Z/A ratio. A generic detector case with an exposure
  time of 30 MTyr, an energy resolution of 20\%, and an angular
  resolution of $0.25 \times (\rm{E/GeV})^{-0.5}$ is shown.  We
  estimated the confidence level using three different density
  models. The solid~(red), dashed~(green), and dotted~(blue) lines
  represent the modified PREM, AK135, and PEM-A, respectively.}
  \label{supfig3}
\end{figure}

\end{document}